\documentclass[a4paper]{jpconf}

\usepackage{citesort}
\usepackage{float}
\usepackage{graphicx}

\begin{document}
\nocite{*}

\title{Short-range correlations for $0\nu\beta \beta$ decay and low-momentum
  $NN$ potentials}

\author{L Coraggio$^1$, N Itaco$^{1,2}$, and R Mancino$^{1,2}$}
\address{$^1$Istituto Nazionale di Fisica Nucleare, \\
Complesso Universitario di Monte  S. Angelo, Via Cintia - I-80126 Napoli, Italy}
\address{$^2$Dipartimento di Matematica e Fisica, Universit\`a degli
  Studi della Campania ``Luigi Vanvitelli'', viale Abramo Lincoln 5 -
  I-81100 Caserta, Italy}

\ead{luigi.coraggio@na.infn.it}

\begin{abstract}
We approach the calculation of the nuclear matrix element of the
neutrinoless double-$\beta$ decay process, considering the
light-neutrino-exchange channel, by way of the realistic shell-model.
In particular the focus of our work is spotted on the role of the
short-range correlations, which should be taken into account because
of the short-range repulsion of the realistic potentials.
Our shell-model wave functions are calculated using an effective
Hamiltonian derived from the high-precision CD-Bonn nucleon-nucleon
potential, the latter renormalized by way of the so-called $V_{\rm
  low\mbox{-}k}$ approach.
The renormalization procedure decouples the repulsive high-momentum
component of the potential from the low-momentum ones by the
introduction of a cutoff $\Lambda$, and is employed to renormalize
consistently the two-body neutrino potentials to calculate the nuclear
matrix elements of candidates to this decay process in 
mass interval ranging from $A=76$ up to $A=136$.
We study the dependence of the decay operator on the choice of the
cutoff, and compare our results with other approaches that can be
found in present literature.
\end{abstract}

\section{Introduction}
\label{intro}
The neutrinoless double-$\beta$ decay is currently one of the main
targets to explore the limits of Standard Model and to understand the
intrinsic nature of the neutrino (see Ref.\cite{Henning16} for a brief
but upgraded review of current and future experiments).
As is well known the detection of such a rare decay would assess the
neutrino as a Majorana particle, namely that neutrinos are their own
anti-particles, and correspond to a lepton number violation, which
will introduce us to ``new physics'' beyond the Standard Model.

On the other side, the measurement of the decay half life may provide
an estimation of its effective mass via the relationship
\begin{equation}
\left[ T^{0\nu}_{1/2} \right]^{-1} = G^{0\nu} \left| M^{0\nu}
\right|^2 \langle m _{\nu} \rangle^2 ~~,
\label{halflife}
\end{equation}
\noindent
where $G^{0\nu}$ is the so-called phase-space factor (or kinematic
factor), $\langle m _{\nu} \rangle$ is the effective neutrino
mass that takes into account the neutrino parameters associated
with the mechanisms of light- and heavy-neutrino exchange, and
$M^{0\nu}$ is the nuclear matrix element (NME) directly related to the
wave functions of the parent and grand-daughter nuclei.

The expression (\ref{halflife}) evidences that a reliable estimate of
the NME is a key point both to understand which are the
most favorable nuclides to detect $0\nu\beta\beta$ decay, and to link
the experimental results to the value of neutrino effective mass.

Currently, the nuclear structure models which are largely employed to
study the $0\nu\beta\beta$ decay of nuclei of experimental interest
are the Interacting Boson Model (IBM) \cite{Barea09,Barea12,Barea13},
the Quasiparticle Random-Phase Approximation (QRPA)
\cite{Simkovic09,Fang11,Faessler12}, Energy Density
Functional methods \cite{Rodriguez10}, and the Shell Model (SM)
\cite{Menendez09a,Menendez09b,Horoi13b,Neacsu15,Brown15}.

One of the issues to be tackled in the calculation of the
$0\nu\beta\beta$ NME is the evaluation of the short-range correlations
(SRC), which account the fact that the action of a two-body decay
operator on an unperturbed (uncorrelated) wave function is not equal
to the action of the same operator on the real (correlated) nuclear
wave function \cite{Bethe71,Kortelainen07}.

This follows from the highly repulsive nature of the nuclear
interaction in its short range, which requires - for nuclear structure
calculations - a consistent regularization of the nucleon-nucleon
($NN$) potential $V^{NN}$ and of any two-body transition operators \cite{Wu85}.

The most common way to soften the matrix elements of the
$0\nu\beta\beta$ decay operator and include SRC is by way of Jastrow
type functions \cite{Miller76,Neacsu12}, and in recent years SRC have
been modeled by the so-called Unitary Correlation Operator Method
(UCOM) \cite{Kortelainen07,Menendez09b}, this approach allowing to
provide a unitary operator which prevents the overlap between the
wave functions of a pair of nucleons \cite{Feldmeier98}.

In this work we present an original approach to the evaluation of SRC
that is consistently linked to the derivation of the effective
shell-model Hamiltonian $H_{\rm eff}$, the latter being calculated
starting from a realistic $NN$ potential.
More precisely, our first step is to consider the high-precision
CD-Bonn $NN$ potential \cite{Machleidt01b}, whose repulsive
high-momentum components are renormalized by way of the $V_{{\rm
    low}\mbox{-}k}$ approach in order to make it suitable for the
derivation of $H_{\rm eff}$ by way of the many-body perturbation
theory \cite{Kuo71,Coraggio12a}.

The renormalization of $V^{NN}$ by way of the $V_{{\rm low}\mbox{-}k}$
procedure \cite{Bogner01,Coraggio09a} occurs through a unitary
transformation $\Omega_{{\rm low}\mbox{-}k}$ in the momentum space of the
two-nucleon Hamiltonian $H^{NN}$, by truncating the full Hilbert space
to a subspace where only relative momenta below a cutoff $\Lambda$ are
allowed. 
Obviously, the unitary transformation preserves the physics of
$H^{NN}$, namely the calculated values of all observables are the same
as those reproduced by the original realistic potential.

This renormalization procedure needs to be applied to any two-body
operator, for consistency reason, before employing the same operator
in nuclear structure calculations which employ wave functions obtained
starting from the same $V_{{\rm low}\mbox{-}k}$.
Consequently, we have renormalized $0\nu\beta\beta$ decay operator by
way of $\Omega_{{\rm low}\mbox{-}k}$ in order to consider effectively the
high-momentum (short range) components of the $NN$ potential, in a
framework where their direct contribution is dumped by the
introduction of a cutoff $\Lambda$.

In the following section we will sketch out a few details of our
theoretical framework, more precisely how the renormalization
procedure of the  $0\nu\beta\beta$ decay operator is carried out.
In Section \ref{results} the results of the calculation of $M^{0\nu}$
for the $0\nu\beta\beta$ decay, within the light-neutrino exchange, of
$^{76}$Ge, $^{82}$Se, $^{130}$Te, and $^{136}$Xe are reported,
comparing those obtained with the bare operator with the results
provided by the renormalization of the decay operator using two
different values of the cutoff $\Lambda$.
The wave functions of the parent and grand-daughter nuclei have been
calculated using the SM $H_{\rm eff}$ reported in
Refs. \cite{Coraggio17a,Coraggio19a}, where the abilities of these
$H_{\rm eff}$ to reproduce the spectroscopic properties of the nuclei
involved in the decay have been extensively reported, as well as the
results of the calculations of two-neutrino double-beta decay NME
$M^{2\nu}$.
Conclusions and perspectives of our work will be reported in Section
\ref{conclusions}.

\section{Theoretical framework}
\label{outline}
\begin{figure}[H]
\begin{center}
\includegraphics[scale=0.7,angle=0]{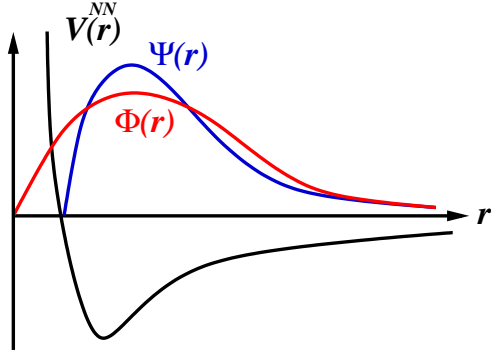}
\end{center}
\caption{Representation of a realistic potential $V^{NN}$, as a
  function of the internucleon distance $r$, and of the correlated and
  non-correlated wave functions $\Psi$ (blue line) and $\Phi$ (red
  line), respectively (see text for details).}
\label{psi}
\end{figure}

As is well known, in nuclear structure calculations with realistic
potentials, practitioners have to face the problem that the basis
states, which constitute the Slater determinants of unperturbed
non-correlated wave functions $\Phi$, are non-zero in the region of
short-range interaction.
This contrasts with the need that, because of the short-range
repulsion of $V^{NN}$ (repulsive high-momentum components in the
momentum space),  the ``real'' correlated wave function $\Psi$ has to
approach to zero as the internucleon distance diminishes, as fast as
the core repulsion increases (see Fig. (\ref{psi})).
This leads to the need to renormalize the short-range (high-momentum)
components of the $NN$ potential, when a perturbative approach to the
many-body problem is pursued.

Here, we briefly introduce the so-called $V_{{\rm low}\mbox{-}k}$ approach,
whose details may be found in Refs. \cite{Bogner01,Coraggio09a}

The eigenvalue problem of the two-nucleon Hamiltonian $H^{NN} (k,k')
=H_0(k,k') + V^{NN} (k,k')$ - $H_0(k,k')$ being the kinetic-energy
term - may be written in the full momentum space of the plane-waves
basis $\langle k | \Psi_{\nu} \rangle$ in the following form:
\begin{center}
\begin{equation}
\int^{\infty}_0 [ H_0(k,k') + V^{NN} (k,k')] \langle k | \Psi_{\nu}
\rangle k^2 dk = E_{\nu} \langle k' | \Psi_{\nu} \rangle~~.
\end{equation}
\label{fullh}
\end{center}
We look for a Hamiltonian $H_{{\rm low}\mbox{-}k} (k,k') = H_0(k,k') +
V_{{\rm low}\mbox{-}k} (k,k')$ that is defined in a reduced subspace
$\displaystyle  P=\int^{\Lambda}_0 | k \rangle \langle k| k^2 dk $,
whose subset of eigenvalues $\{\tilde{E}_{\mu}\}_{\mu \in P}$ belongs
to the set of eigenvalues $\{E_{\nu}\}$ of the Hamiltonian
$H^{NN}(k,k')$ defined in full Hilbert space:
\begin{equation}
\int^{\Lambda}_0 [ H_0(k,k') + V_{{\rm low}\mbox{-}k} (k,k')] \langle
k | \Phi_{\mu} \rangle k^2 dk =\tilde {E_{\mu}} \langle k' |
\Phi_{\mu} \rangle ~~.
\end{equation}
\label{effh}
This goal may be achieved through a similarity transformation
$\Omega_{{\rm low}\mbox{-}k}$, that leads to the identity $\mathcal{H}
  = \Omega_{{\rm low}\mbox{-}k}^{-1} H^{NN} \Omega_{{\rm low}\mbox{-}k}$.
$\Omega_{{\rm low}\mbox{-}k}$ needs to satisfy the decoupling
condition which decouples the low-momentum subspace $P$ from its
complement $Q=1-P$:
\begin{equation}
 Q \mathcal{H} P = Q \Omega^{-1}_{{\rm low}\mbox{-}k} H^{NN}
   \Omega_{{\rm low}\mbox{-}k} P = 0
\end{equation}
\label{decoup}
A very convenient expression of the operator $\Omega_{{\rm
    low}\mbox{-}k}$  may be obtained according to the Lee-Suzuki
  formulation \cite{Suzuki80}, which is:
\begin{eqnarray}
\label{leesuzuki}
 \Omega_{{\rm low}\mbox{-}k} P = I_P &~~~~& P \Omega_{{\rm low}\mbox{-}k} Q = 0 \\
 Q \Omega_{{\rm low}\mbox{-}k} P = \omega &~~~& Q \Omega_{{\rm
                                                low}\mbox{-}k} Q = I_Q
                                                ~~,\nonumber
\end{eqnarray}
\noindent
were $I_P,I_Q$ represents the identity operator in the $P$ and $Q$
spaces, respectively.
This form leads to a non-linear matrix equation for the $\omega$
operator, which can be solved using iterative techniques
\cite{Andreozzi96}:
\begin{equation}
 Q H^{NN} P + Q H^{NN} Q \omega - \omega P H^{NN} P - \omega P H^{NN} Q \omega = 0 
 ~~.
\label{ls}
\end{equation}
Once Eq. (\ref{ls}) is solved and the operator $\omega$ is obtained, an
hermitization procedure, based on the Cholesky decomposition of the
operator $\Omega_{{\rm low}\mbox{-}k}$ \cite{Andreozzi96}, evolves the
Lee-Suzuki similarity transformation to a unitary transformation.

The $V_{{\rm low}\mbox{-}k}$, which is explicitly zero for momenta
above the cutoff $\Lambda$, may now be suitable as an input for the
derivation of $H_{\rm eff}$ by way of the many-body perturbation
theory \cite{Coraggio12a}.
In Refs. \cite{Coraggio17a,Coraggio19a} we have reported the results
for the calculation of $M^{2\nu}$ for $^{76}$Ge, $^{82}$Se,
$^{130}$Te, and $^{136}$Xe, and in the present work we employ the same
wave functions obtained from the diagonalization of $H_{\rm eff}$s
obtained renormalizing the CD-Bonn potential with a cutoff
$\Lambda=2.6$ fm$^{-1}$.

As regards the calculation of $M_{\alpha}^{0\nu}$- $\alpha$ denoting
the Fermi ($F$), Gamow-Teller (GT), or tensor ($T$) decay channels -
we recall that, within the closure approximation \cite{Senkov14}, it
can be written in terms of the two-body transition-density matrix
elements $\langle f | a^{\dagger}_{p}a_{n} a^{\dagger}_{p^\prime}
a_{n^\prime} | i \rangle$, the indices $i,f$ denoting the parent and
grand-daughter nuclei:
\begin{equation}
\label{M0nu}
  M_\alpha^{0\nu}=  \sum_{j_n j_{n^\prime} j_p j_{p^\prime} J_\pi}
  \langle f | a^{\dagger}_{p}a_{n} a^{\dagger}_{p^\prime} a_{n^\prime}
  | i \rangle \left< j_p  j_{p^\prime}; J^\pi \mid \tau^-_{1}
    \tau^-_{2}O^\alpha_{12} \mid  j_n j_{n^\prime} ; J^\pi \right>~~.
\end{equation}
The operators $O^{\alpha}_{12}$ are expressed in terms of the
neutrino potentials $H^{\alpha}$ and form functions $h^{\alpha}(q)$:
\begin{eqnarray} 
 O_{12}^{GT} & = & \vec{\sigma}_1 \cdot \vec{\sigma}_2 H_{GT}(r) \\
\nonumber O_{12}^{F} & = & H_{F}(r)  \\ 
\nonumber O_{12}^{T} & = & \left[3\left(\vec{\sigma}_1 \cdot \hat{r} \right) 
\left(\vec{\sigma}_1 \cdot \hat{r} \right)  \vec{\sigma}_1 \cdot
                           \vec{\sigma}_2 \right] H_{T}(r)~~,
\label{operator}
\end{eqnarray}
\begin{equation}
H_{\alpha}(r)=\frac {2R}{\pi} \int_{0}^{\infty} \frac {j_{n_{\alpha}}(qr)
  h_{\alpha}(q^2)qdq}{q+\left< E \right>}~~.
\label{neutpot}
\end{equation}
\noindent
The average energies $\left< E \right>$ have been evaluated as in
Ref. \cite{Haxton84}, the parameter $R$ is $R=1.2 A^{1/3}$ fm, and
$j_{n_{\alpha}}(qr)$ are the spherical Bessel functions, $n_{\alpha}=0$
for Fermi and Gamow-Teller components, $n_{\alpha}=2$ for
the tensor one.
The explicit expression of neutrino form functions $h_{\alpha}(q)$ for
light-neutrino exchange may be found in Ref. \cite{Horoi10}.

We transform the operators in Eq. (\ref{operator}), that are expressed
in a local configuration-space form, to a momentum-space
representation \cite{Veerasamy11}, in order to construct a
low-momentum decay operator $O^{\alpha}_{{\rm low}\mbox{-}k} =
P\Omega^{-1}_{{\rm low}\mbox{-}k} O^{\alpha} \Omega_{{\rm
    low}\mbox{-}k}P$.
The latter unitary transformation allows to take into account
effectively the high-momentum (short-range) correlations on the
two-nucleon wave function.
In the following Section, the results of the calculation of
$M_{\alpha}^{0\nu}$ for $^{76}$Ge, $^{82}$Se, $^{130}$Te, and
$^{136}$Xe decays will be presented, using both bare and renormalized
$0 \nu \beta \beta$ decay operators.

\section{Results}
\label{results}
Our starting point is the calculation of nuclear wave functions of
parent and grand-daughter nuclei for $^{76}$Ge, $^{82}$Se, $^{130}$Te, and
$^{136}$Xe decays, using the SM effective Hamiltonians derived from
CD-Bonn potential, the latter being renormalized via the $V_{{\rm
    low}\mbox{-}k}$ procedure employing a cutoff $\Lambda=2.6$
fm$^{-1}$ \cite{Coraggio17a,Coraggio19a}.
In those works, where tables with the theoretical single-particle
energies and two-body matrix elements of the residual interaction have
been also reported, it has been carried out an extensive study of the
doubly-beta decay of such nuclei, with the perspective to check the
theoretical framework for future studies of their $0 \nu \beta \beta$
decay.

In Table (\ref{results00}) the results of the calculations of
$M^{0\nu}$ for the $^{76}$Ge, $^{82}$Se, $^{130}$Te, and $^{136}$Xe
decays are reported.
We have neglected the tensor component in the expression (\ref{M0nu}),
since its contribution is about 2-3 order of magnitude smaller of the
Fermi and Gamow-Teller components.

The calculations have been performed both with bare $0 \nu \beta
\beta$ operators and those renormalized via the $V_{{\rm
    low}\mbox{-}k}$ approach, as reported in Section \ref{outline}.
We have employed for the renormalization of the high-momentum
components of the decay operator two cutoffs, $\Lambda=2.6,2.1$
fm$^{-1}$, in order to evaluate the dependence of the results on this
choice.

\begin{center}
\begin{table}[h]
\label{results00}
\caption{Results for the calculation of $M^{0\nu}$ relative to
  $^{76}$Ge, $^{82}$Se, $^{130}$Te, and $^{136}$Xe
  decays. Calculations with bare operator are compared with results
  obtained taking into account the SRC by way of the $V_{{\rm
      low}\mbox{-}k}$ approach and cutoffs $\Lambda=2.6,2.1$
  fm$^{-1}$. In parentheses they are reported the variations in
  percentage of SRC results with respect to the bare ones.} 
\centering
\begin{tabular}{ccccc}
\br
 Decay & bare operator & $\Lambda=2.6$ fm$^{-1}$& $\Lambda=2.1$ fm$^{-1}$ \\
\mr
 $^{76}$Ge $\rightarrow$ $^{76}$Se & 3.35 & 3.29 $(1.8\%)$ & 3.27 $(2.4\%)$ \\ 
 $^{82}$Se $\rightarrow$ $^{82}$Kr & 3.30 &  3.25 $(1.5\%)$ & 3.23 $(2.1\%)$ \\ 
 $^{130}$Te $\rightarrow$ $^{130}$Xe & 3.27 &  3.22 $(1.6\%)$ & 3.20 $(2.1\%)$ \\ 
 $^{136}$Xe $\rightarrow$ $^{136}$Ba & 2.47 &  2.43 $(1.6\%)$ & 2.41 $(2.4\%)$ \\ 
\br
\end{tabular}
\end{table}
\end{center}

As can be seen, the variation of the calculated $M^{0\nu}$, with
respect to the ones obtained with the bare operator, is about $2\%$,
the softening being mildly larger with the smaller cutoff
$\Lambda=2.1$ fm$^{-1}$ since it corresponds to a larger
renormalization effect.
The result that the $\Omega_{{\rm low}\mbox{-}k}$ transformation leads
to a tiny renormalization effect can be ascribed to the behavior of
neutrino form functions $h^{\alpha}(q)$, which approach rapidly to zero for
momenta $q\rightarrow \infty$ \cite{Horoi10}.
Consequently, they are scarcely sensitive to the renormalization of
high-momentum components by the $\Omega_{{\rm low}\mbox{-}k}$
operator.

It should pointed out that the effect in magnitude of this
renormalization is very close to the one obtained by way of UCOM SRC
by Menendez and coworkers (see Table 8 in Ref. \cite{Menendez09b})
for the same nuclear decays, leading to a lighter softening of NME
with respect to the one provided by Jastrow type SRC.
As a matter of fact, the authors experienced a reduction of the
calculated NMEs, with respect the bare decay operator, about 20-25$\%$
employing standard Jastrow type correlations, and 5-6$\%$ the UCOM
ones.

It is worth to stress again that our calculations manage the correlations
induced by the renormalization of the high-momentum components of
$V^{NN}$ on an equal footing, both for the $NN$ potential and the
two-body matrix elements of the $0 \nu \beta \beta$ decay operator.
It should be also mentioned that a similar approach was pursued in
works by Kuo and coworkers \cite{Wu85,Song91}, where the
renormalization of realistic potentials by way of the reaction matrix
$G$ was employed to calculate SRC in terms of the defect wave
functions \cite{Bethe71}.

\section{Conclusions and perspectives}
\label{conclusions}
In this work we have introduced an original approach to consider the
effects of short-range correlations in the calculation of the nuclear
matrix element for the $0 \nu \beta \beta$ decay within the realistic
shell model.

This has been done by renormalizing the two-nucleon Hamiltonian for a
realistic $NN$ potential and the $0 \nu \beta \beta$ decay operator,
consistently, by way of the so-called $V_{{\rm low}\mbox{-}k}$
approach \cite{Bogner01}.
Then, we have  calculated the $M^{0 \nu}$ for $0 \nu \beta
\beta$-decay candidates $^{76}$Ge, $^{82}$Se, $^{130}$Te, and
$^{136}$Xe, using both the bare decay operator and the renormalized
one.
The wave functions of parent and grand-daughter nuclei employed for
these calculations are the same as in
Refs. \cite{Coraggio17a,Coraggio19a}, where the SM effective
Hamiltonians have been derived by way of the many-body
perturbation theory from the CD-Bonn $NN$ potential, the latter being
renormalized  employing the $V_{{\rm low}\mbox{-}k}$ method.

Our results show that this novel approach to the evaluation of SRC
reveals a tiny effect, when compared to the inclusion of standard
Jastrow type correlations.

The next step will be to build up SM effective operators for the
two-body $0 \nu \beta \beta$-decay operator by way of the many-body
perturbation theory, as in Refs. \cite{Wu85,Holt13d}, consistently with
the derivation of the SM Hamiltonian from realistic $NN$ potentials.

Our goal is to perform fully-consistent calculations of $M^{0 \nu}$
which avoids to resort to parameters fitted to experiment, providing
an improvement of the reliability and predictivity of
nuclear-structure calculations for the $0 \nu \beta \beta$ decay.

\section*{References}
\bibliography{biblio}

\end{document}